\documentclass[%
 aip,
 jmp,%
 amsmath,amssymb,
 reprint,%
]{revtex4-1}

\usepackage{graphicx}
\usepackage{dcolumn}
\usepackage{bm}
\usepackage{caption}
\usepackage{blindtext}
\usepackage{float}
\usepackage{caption}
\usepackage{subfig}
\usepackage{afterpage}
\usepackage{adjustbox}

\begin{document}

\preprint{AIP/123-QED}

\title{Optimisation of the pointing stability of laser-wakefield accelerated electron beams}
\author{R. J. Garland}%
\affiliation{School of Mathematics and Physics, Queens University Belfast, University Road, Belfast, BT7 1NN, Northern Ireland}
\author{K. Poder}
\affiliation{Imperial College of Science, Technology and Medicine, London SW7 2BZ, UK}
\author{J. Cole}
\affiliation{Imperial College of Science, Technology and Medicine, London SW7 2BZ, UK}
\author{W. Schumaker}
\affiliation{Center for Ultrafast Optical Science, University of Michigan, Ann Arbor, Michigan 48109-2099, USA}
\author{D. Doria}
\affiliation{School of Mathematics and Physics, Queens University Belfast, University Road, Belfast, BT7 1NN, Northern Ireland}
\author{L. A. Gizzi}
\affiliation{Istituto Nazionale di Ottica, Consiglio Nazionale delle Ricerche, 56124 Pisa, Italy}
\affiliation{INFN, Sez. Pisa, Largo B. Pontecorvo, 3-56127 Pisa, Italy}
\author{G. Grittani}
\affiliation{Istituto Nazionale di Ottica, Consiglio Nazionale delle Ricerche, 56124 Pisa, Italy}
\affiliation{INFN, Sez. Pisa, Largo B. Pontecorvo, 3-56127 Pisa, Italy}
\author{K. Krushelnick}
\affiliation{Center for Ultrafast Optical Science, University of Michigan, Ann Arbor, Michigan 48109-2099, USA}
\author{S. Kuschel}
\affiliation{Helmholtz Institute Jena, Fr\"{o}belstieg 3, 07743 Jena, Germany}
\author{S. P. D. Mangles}
\affiliation{Imperial College of Science, Technology and Medicine, London SW7 2BZ, UK}
\author{Z. Najmudin}
\affiliation{Imperial College of Science, Technology and Medicine, London SW7 2BZ, UK}
\author{D. Symes}
\affiliation{Central Laser Facility, Rutherford Appleton Laboratory, Didcot, Oxfordshire OX11 0QX, United Kingdom}
\author{A. G. R. Thomas}
\affiliation{Center for Ultrafast Optical Science, University of Michigan, Ann Arbor, Michigan 48109-2099, USA}
\author{M. Vargas}
\affiliation{Center for Ultrafast Optical Science, University of Michigan, Ann Arbor, Michigan 48109-2099, USA}
\author{M. Zepf}
\affiliation{School of Mathematics and Physics, Queens University Belfast, University Road, Belfast, BT7 1NN, Northern Ireland}
\affiliation{Helmholtz Institute Jena, Fr\"{o}belstieg 3, 07743 Jena, Germany}
\author{G. Sarri}
\email{g.sarri@qub.ac.uk}
\affiliation{School of Mathematics and Physics, Queens University Belfast, University Road, Belfast, BT7 1NN, Northern Ireland}

\date{\today}

\begin{abstract}
Laser-wakefield acceleration is a promising technique for the next generation of ultra-compact, high-energy particle accelerators. However, for a meaningful use of laser-driven particle beams it is necessary that they present a high degree of pointing stability in order to be injected into transport lines and further acceleration stages. Here we show a comprehensive experimental study of the main factors limiting the pointing stability of laser-wakefield accelerated electron beams. It is shown that gas-cells provide a much more stable electron generation axis, if compared to gas-jet targets, virtually regardless of the gas density used. A sub-mrad shot-to-shot fluctuation in pointing is measured and a consistent non-zero offset of the electron axis in respect to the laser propagation axis is found to be solely related to a residual angular dispersion introduced by the laser compression system and can be used as a precise diagnostic tool for compression oprtimisation in chirped pulse amplified lasers. 
\end{abstract}

\pacs{Valid PACS appear here}
\maketitle


\section{Introduction}

\indent Laser Wakefield acceleration (LWFA) exploits the excitation of quasi-electrostatic plasma waves during the propagation of an intense laser pulse through an underdense plasma. \cite{Tajima,Esarey,Popp,Vargas,Pollock,Leemans2006,Wang} The intense accelerating fields that can be achieved ($\approx$ 10s of GV/m, compared to 10 - 100 MV/m in solid-state conventional accelerators) are at the core of important practical applications such as generation of secondary x-ray and gamma-ray radiation \cite{Corde1,Corde2,Puhoc,DiPiazza} and are indeed encouraging towards the construction of ultra-compact high-energy particle accelerators.

\indent Efficient guiding of the laser is necessary in order to ensure acceleration over distances exceeding the laser's Rayleigh range and mm- to cm-scale acceleration is now routinely achieved by using two main classes of gas targets: gas-jets and gas-cells. Gas-jets were first to be developed and operate via up-shooting a supersonic flow of gas from a nozzle. This gas delivery system presents significant limitations in providing a stable gas target: it presents sharp pressure gradients at the gas-vacuum interface, and it excites internal shocks that induce local non-uniformities in the gas density.\cite{Esarey} These limitations are overcome by gas-cell targets, which instead are pre-filled with gas before the interaction with the laser. \cite{Popp, Vargas}

\indent However, experimental and theoretical work has demonstrated that LWFA is affected by significant shot-to-shot fluctuations \cite{Vargas,Popp} in the electron beam pointing and a series of reasons have been identified. These might include: spatial non-uniformity of the laser intensity profile and phase front, mechanical vibrations of the optical components, non-uniformities in the gas density profile, and strong non-linearities in the laser-plasma coupling. These fluctuation might prove detrimental for transport of the beam or for injection in further acceleration stages. \cite{Blumenfeld}

\indent In this paper we report on a detailed study of the main parameters affecting the stability of LWFA electron beams. Confirming other works reported in the literature, we experimentally show that a gas-cell target provides a much smaller shot-to-shot fluctuation in the electron beam pointing if compared to gas-jet targets with equal electron density. Moreover, the pointing stability appears to be virtually independent from the plasma density used, in a regime whereby efficient electron injection is still occurring, suggesting that laser-plasma coupling can be neglected when considering this problem. A consistent offset between the electron beam and laser propagation axis is also measured, and it is found to be directly related to a phase front tilt in the laser beam. In agreement with recent experimental work, \cite{Popp}, we find the laser front tilt directly correlated to a residual angular dispersion introduced by a non perfect alignment of the compressor gratings in the laser chain. Fine tuning of the latter is able to eliminate this offset, proposing this effect as an accurate diagnostic tool for the compression system of ultra-short lasers. Finally we show that, by careful control of all these parameters, it is possible to generated electron beams with sub-mrad pointing fluctuation and virtually exactly parallel to the laser propagation axis.

\indent The structure of the paper is as follows: in Section 2 a description of the experimental setup will be given whereas the main experimental results will be presented in Sections 3 and 4. Finally, a conclusive paragraph will be given in Section 5. 

\section{Experimental Setup}

\begin{figure}[h]
	\begin{center}
\includegraphics[width=0.5\textwidth]{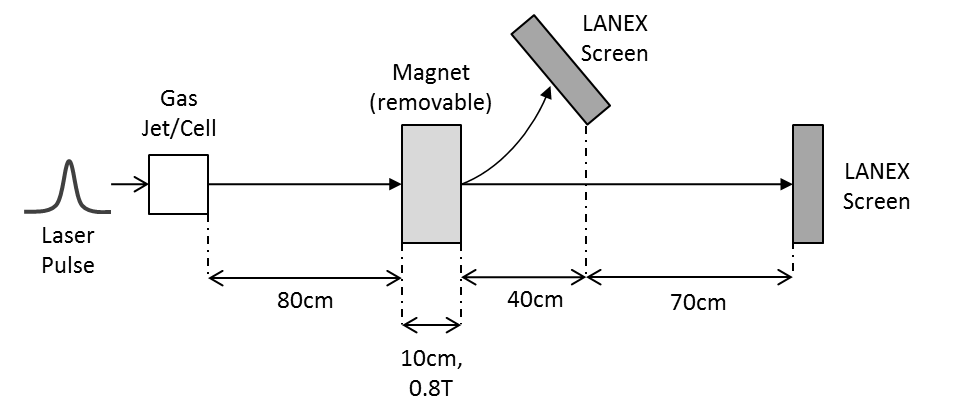} 
		\caption{Sketch of the experimental setup}
		\label{fig:fig1}
	\end{center}
\end{figure}

\indent The experiment was carried out using the Astra-Gemini Laser at the Rutherford Appleton Laboratory \cite{Gemini}, which delivered a laser beam with a central wavelength of 0.8$\mu$m and a pulse duration of 42 $\pm$ 4 fs. This beam was focussed using a f/20 off axis parabola down to a focal spot with a full width at half maximum of $\sim$ (27 $\pm$ 3) $\mu$ containing 50\% of the initial laser energy (14 J, peak intensity of $\sim$ 2$\times$10$^{19}$W/cm$^{2}$). The laser was linearly polarised in a horizontal direction, i.e. perpendicularly to the main axis of gas outflow from the gas-jet. Figure 1 shows the setup used during the experiment. 
\begin{figure}[h]
	\begin{center}
\includegraphics[width=0.5\textwidth]{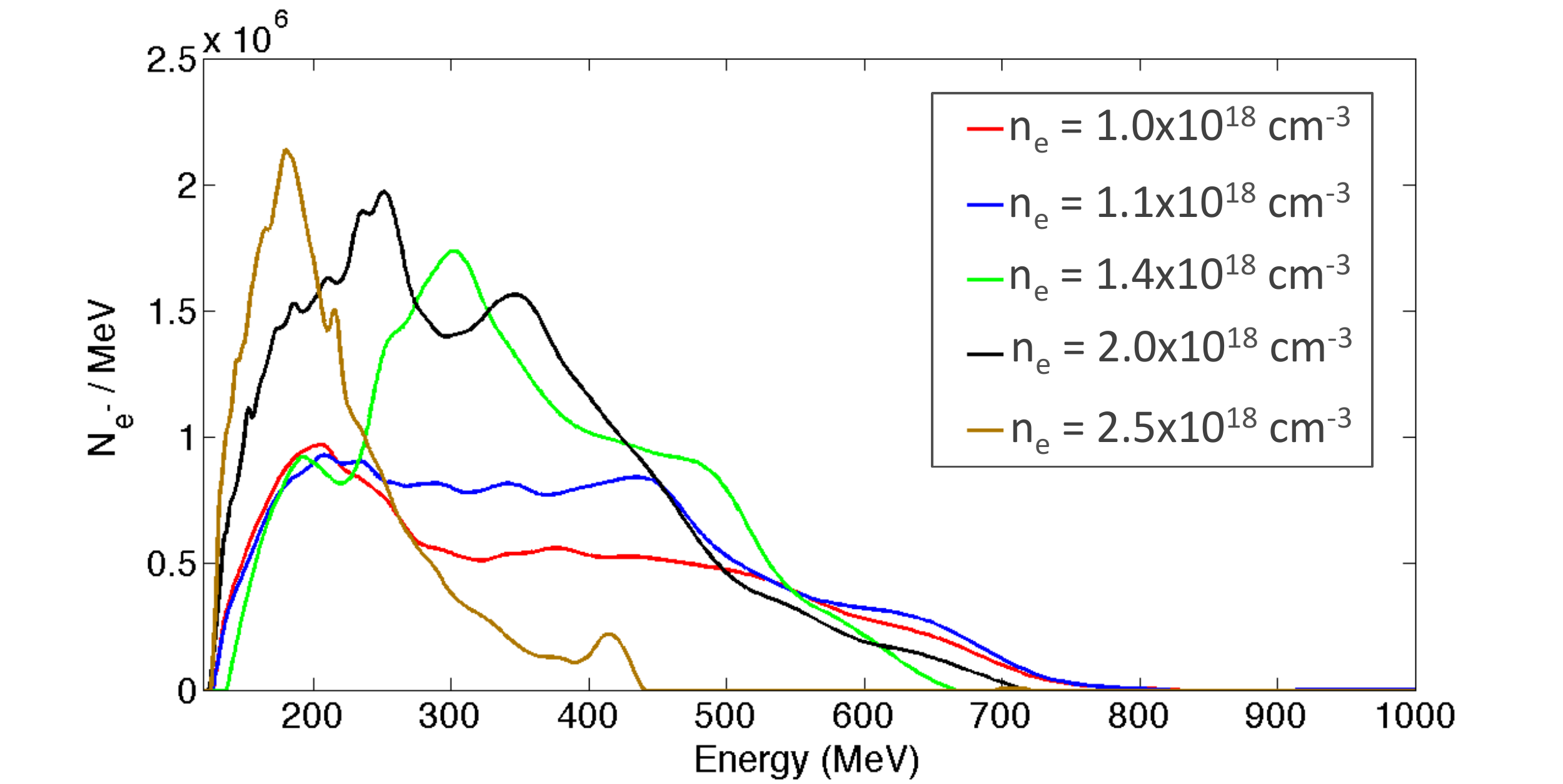} 
		\caption{Typical electron spectra obtained during the experiment, for different electron densities.}
		\label{fig:fig1}
	\end{center}
\end{figure}

\indent The gas target was provided by either a gas-cell or a gas-jet and in both cases He with a 1\% doping of N$_2$ was used, in order to operate in an ionisation injection scheme \cite{Esarey,ionization} 15mm-long gas-cell were filled with a gas pressure ranging from 200 to 1000 mbar whereas the gas-jet operated at a backing pressure between 45 and 55 bar. Optical interferometry of the resulting laser-generated plasma indicated an electron density of 1 - 5 $\times10^{18}$ cm$^{-3}$ for the gas-cell and 2 $\times10^{18}$ cm$^{-3}$ for the gas-jet. The main diagnostics used for the experiment were a particle magnetic spectrometer and an on-axis profile imager. The first consisted of a 0.8 T, 10cm long pair of magnets and an off-axis LANEX screen \cite{LANEX}. The spectrometer could resolve electron energies between 120 MeV and 1.2 GeV. The spectrometer was cross-calibrated using absolutely calibrated Imaging Plates \cite{IP}. The profile imager consisted of a LANEX screen placed on the laser propagation axis (distances detailed in Fig. 1). This scintillator was placed 2m away from the exit of the gas target and was imaged by a CCD camera with a 10x magnification, implying a sub-mrad resolution of the electron beam pointing. 

\indent Fig. 2 shows the typical electron spectra obtained during the experiment for different gas densities. All spectra present a broad spectrum, extending up to 600 - 700 MeV with a typical charge of the order of hundreds of pC. Stable electron acceleration is achieved in a density window of 1 to 2.5 $\times10^{18}$ cm$^{-3}$ with the latter presenting a much lower maximum energy compared to the others. Even higher pressure failed to produce stable electron beams, a clear indication of significant dephasing of the laser pulse through the gas \cite{Esarey}.

\section{Experimental results: gas-cell vs. gas-jets}

\begin{figure}[b]
	\begin{center}
\includegraphics[width=0.5\textwidth]{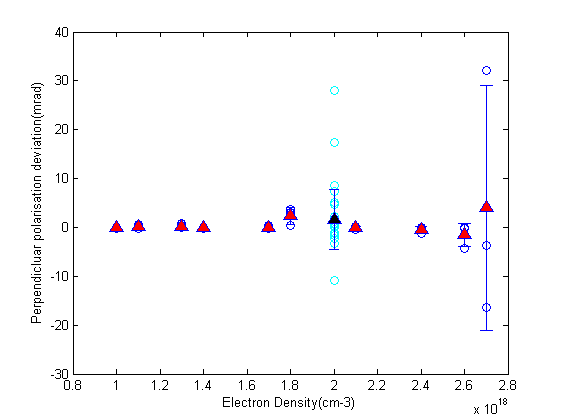} 
		\caption{Pointing of the laser-driven electron beam along the axis perpendicular to the laser propagation axis) for different plasma densities. Light blue (dark blue) circles represent single-shot measurements with a gas-jet (gas-cell) target. Triangles and lines represent average and standard deviation, respectively, for each plasma density.}
		\label{fig:fig1}
	\end{center}
\end{figure}

\begin{figure}[b]
	\begin{center}
\includegraphics[width=0.5\textwidth]{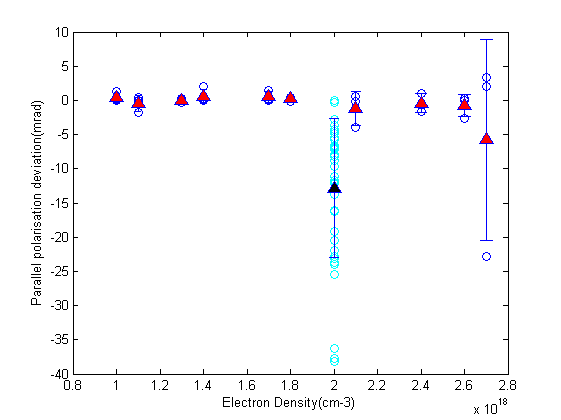} 
		\caption{Pointing of the laser-driven electron beam along the axis parallel to the laser propagation axis for different plasma densities. Light blue (dark blue) circles represent single-shot measurements with a gas-jet (gas-cell) target. Triangles and lines represent average and standard deviation, respectively, for each plasma density.}
		\label{fig:fig1}
	\end{center}
\end{figure}

\indent Figs. 3 and 4 show the measured pointing of the electron beam for different plasma densities, in respect to the laser propagation axis, in the directions both perpendicular (Fig. 3) and parallel (Fig. 4) to the laser polarisation axis. Gas-cell targets are seen to provide a reasonably small shot-to-shot fluctuation in pointing, which is always smaller than 1 mrad for each gas density in which LWFA was efficiently triggered. Only at a density of 2.6$\times10^{18}$cm$^{-3}$, larger fluctuation are observed, but this relatively large gas density proved to be highly unstable in generating electrons. It must be noted that a slight difference is detected for the axis parallel and perpendicular to the laser polarisation axis (0.8 mrad in the parallel case, compared to an average of 0.3 mrad in the opposite axis) but the difference is too small to provide a statistically significant set of data. 

\indent It is worth noticing though that the average angular deviation from the laser propagation axis is zero only for the axis parallel to the laser polarisation but of the order of 1 mrad along the opposite axis. We attribute this deviation to a residual pulse front tilt in the laser beam, and a more detailed characterisation of this phenomenon will be given in the next Section. It is interesting to notice that the pointing fluctuation and deviation from the laser propagation axis appear to be virtually uncorrelated with the plasma density, suggesting that in a density window in which LWFA is efficiently triggered the laser-plasma coupling does not play a significant role in the electron beam pointing. 

\indent Gas-jet targets present instead a much larger shot-to-shot fluctuation in pointing (10 mrad and 6 mrad in the parallel and perpendicular axis, respectively). This can be easily understood if we consider that a gas-jet would present a much less uniform plasma density profile and that non-uniformities would be randomly distributed at each shot, in agreement with what observed in Ref. \cite{Vargas}. It is also interesting to notice that the average deviation from the laser polarisation axis appears to be, in this case, significantly different on the two main axis: whilst it is practically comparable to the gas-cell case perpendicularly to the laser propagation axis ($\approx$ 1.7 mrad), it is much larger along the other axis ($\approx$ 13 mrad). We can explain this by considering that a gas-jet would present a round gas distribution with a very sharp density gradient along the laser propagation axis; when the laser encounters such a sharp gradient, it would tend to propagate through regions that locally have a lower refractive index (and, therefore, a lower plasma density) with the possibility of it being steered from the original axis of propagation. This would of course occur only along the axis perpendicular to the gas outflow and, therefore, parallel to the laser propagation axis. On the other hand, a gas-cell would present a flatter density distribution with a much shallower density gradient along the laser propagation axis. This facilitates laser guiding and avoids beam steering.

\section{Experimental results: pulse front tilt effects}

\begin{figure}[h]
	\begin{center}
\includegraphics[width=0.5\textwidth]{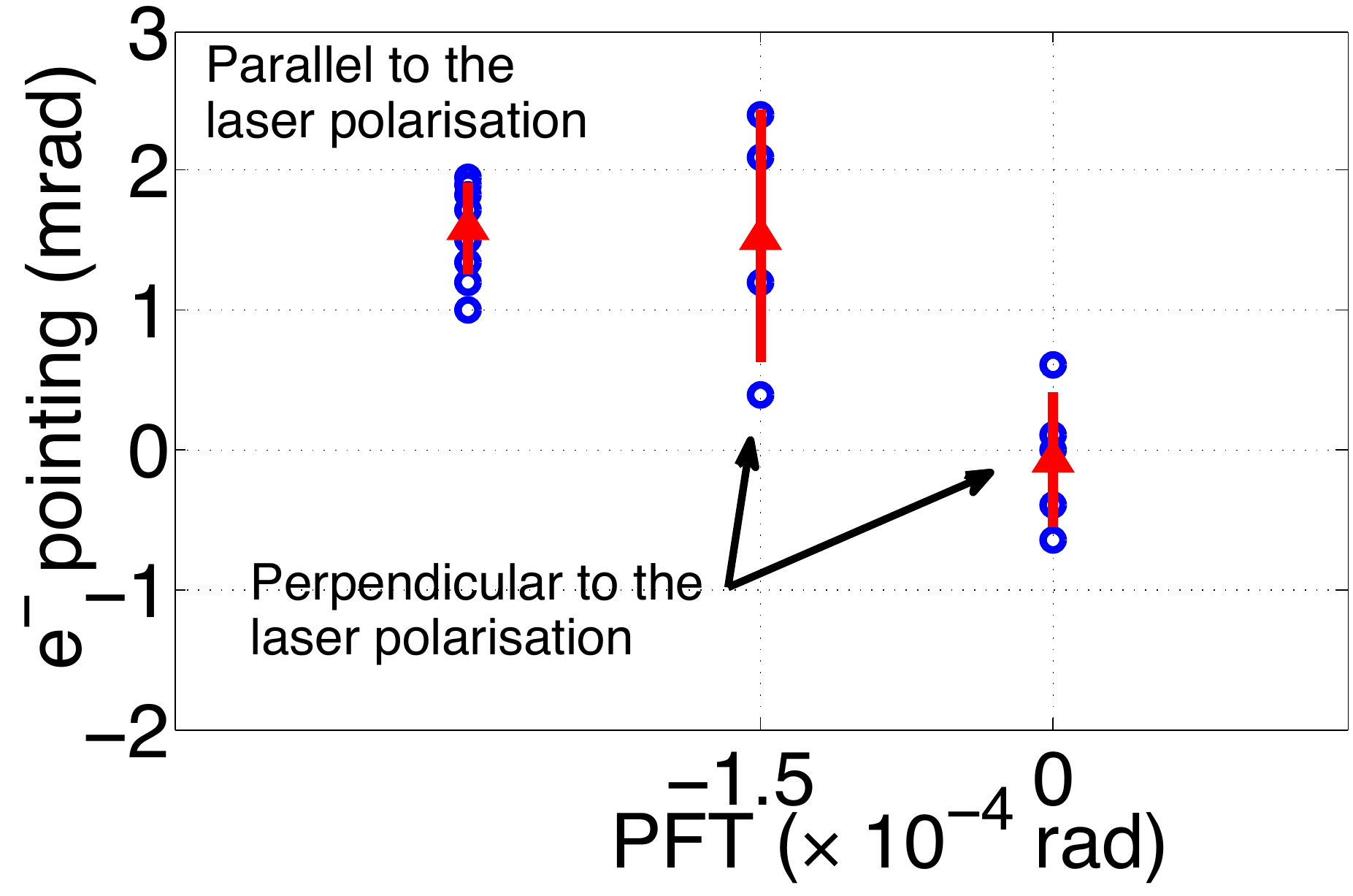} 
		\caption{Effect of the rotation of the compressor gratings on the pointing of the electron beam. Blue circles represent single shot data, whereas triangles and lines represent average and standard deviation, respectively, for each set of data. Data on the left represent the electron beam pointing parallel to the laser propagation axis whereas the two sets of data on the right indicate the beam pointing perpendicular to the laser propagation axis for different laser pulse front tilts.}
		\label{fig:fig1}
	\end{center}
\end{figure}   

\indent The effect of pulse front tilt on the direction of the LWFA electron beams was first experimentally detected by Popp et al. \cite{Popp} and, in order to fully appreciate the results reported here, we will discuss again some of the key theoretical elements. 

\indent In a chirped pulse amplified laser beam\cite{Strickland}, a slight deviation from a perfect parallelism between the compressor diffraction gratings will induce an angular chirp, which will result into a laser intensity profile that is tilted with respect to the laser-propagation direction.\cite{Popp,Akturk,Osvay} Following the notation of Popp et al., if we call $\epsilon$ the relative angle between the two gratings, $s$ the groove spacing, $\beta$ the diffraction angle of  single grating, and $\alpha$ the laser angle of incidence onto the grating, the angular chirp can be estimated as: d$\phi$/d$\lambda = 2\epsilon\tan\beta / (s\cos\alpha)$. The tilt between the laser propagation axis and its phase front will then be related to the angular chirp by: $\tan \psi = \lambda_0$d$\phi$/d$\lambda$, with $\lambda_0$ being the laser central wavelength. Popp et al. \cite{Popp} have shown that a non-zero pulse front tilt will induce a net deviation of the electron beam axis from the laser propagation axis, in an almost linear fashion. 

This suggests that the non-zero deviation from the laser and the electron propagation axis that we observe with the gas-cell, in a direction perpendicular to the laser polarisation axis, might be due to a slight misalignment in the compressor gratings. Indeed, by rotating one of the two compressor gratings by 108 $\mu$rad, this deviation is seen to go down to zero. This is shown in Fig. \ref{}. It must be said that such rotation of the grating will induce a slight temporal stretching of the laser beam of approximately 7.5 fs and that this was corrected using a DAZZLER system. 

\indent In the case of our experiment, $s = 750$ nm, $\beta=30^\circ$, and $\alpha=30^\circ$ and, by taking $\epsilon= 108 \mu$rad, we can estimate the angular chirp to be d$\phi$/d$\lambda\approx 0.17 \mu$rad/nm, implying a pulse front tilt of $\psi\approx1.5\times10^{-4}$ rad. Such a pulse front tilt induces in our experiment a deviation of the electron beam axis from the laser propagation axis of approximately 1 mrad, which is in line with the results reported by Popp et al. \cite{Popp}. This is an interesting result, since it seems to suggest that this effect is quite consistent even to very small angles. As expected, a pulse front tilt has an effect only perpendicularly to the laser polarisation axis and has no effect on the other axis. It is interesting to note that this effect artificially amplifies the angular misalignment of the compressor by approximately one order of magnitude: a 108 $\mu$rad misalignment induces approximately a 1 mrad deviation of beam pointing. This suggests that a careful characterisation of the electron beam pointing can represent an efficient diagnostic tool for fine optimisation of the laser compressor, a necessary pre-requisite for ultra-high intensity, ultra-short laser pulses.

\section{Conclusions}
A study of the pointing stability of laser-wakefield accelerated electron beams has been carried out in order to isolate the main parameters that might affect it. A gas-cell target is seen to provide much more stable electron beams if compared to a gas-jet, clear indication that density non-uniformities in the gas density profile significantly affect the electron beam pointing. However, pointing stability is not seen to change across the plasma densities in which laser-wakefield can be triggered, possibly suggesting that laser-plasma coupling is not significantly influencing the electron beam pointing. Moreover, the non-zero average deviation of the electron beam axis with respect to the laser propagation axis has been found to be extremely sensitive to the degree of parallelism between the compressor gratings of the laser. This high degree of sensitivity promotes measurements of the electron beam pointing as an efficient diagnostic tool for the optimisation of laser compressors in ultra-short laser systems.

\section{Acknowledgements}
The authors are grateful for the technical support of the CLF staff. G. Sarri wishes to acknowledge financial support from EPSRC (grant No.: EP/L013975/1).

\end{document}